\documentclass[conference]{IEEEtran}
\usepackage{graphicx}
\usepackage{subcaption}
\usepackage{amsmath,amssymb,epsfig} 
\usepackage[table]{xcolor}
\usepackage{tabularx,multirow,array}
\usepackage{times}
\usepackage{booktabs}
\usepackage{multirow}
\usepackage{soul}
\usepackage{subcaption}  
\usepackage{enumerate}
\usepackage[countmax]{subfloat}
\usepackage{algpseudocode}
\usepackage{algorithm}
\usepackage{threeparttable}
\usepackage{cite}
\usepackage{url}
\usepackage{amsthm}
\usepackage[utf8x]{inputenc}

\usepackage{nicematrix}
\usepackage{tikz}
\usetikzlibrary{patterns}
\tikzset
  { 
    style a/.style = {pattern = north east lines, pattern color=black!70} ,
    style b/.style = {pattern = north west lines, pattern color=black!70}
  }

\makeatletter
\def\endthebibliography{%
  \def\@noitemerr{\@latex@warning{Empty `thebibliography' environment}}%
  \endlist
}
\makeatother

\begin{document}

\title{On Efficient Topology Management in Service-Oriented 6G Networks: An Edge Video Distribution Case Study}\author{
\IEEEauthorblockN{
Zied Ennaceur\textsuperscript{1}, Mounir Bensalem\textsuperscript{1}, Admela Jukan\textsuperscript{1}, Claus Keuker\textsuperscript{2} \\
Huanzhuo Wu\textsuperscript{3}, Rastin Pries\textsuperscript{3} 
}

\vspace{0.2cm} 

\IEEEauthorblockA{
\textsuperscript{1}Technische Universität Braunschweig, Germany,
\textsuperscript{2}Smart Mobile Labs, Germany,
\textsuperscript{3}Nokia, Germany
}
\IEEEauthorblockA{
\textsuperscript{1}\{zied.ennaceur, mounir.bensalem, a.jukan\}@tu-bs.de,
\textsuperscript{2}\{claus.keuker\}@smartmobilelabs.com\\
\textsuperscript{3}\{huanzhuo.wu, rastin.pries\}@nokia.com
}
}

\providecommand{\keywords}[1]{\textbf{\textit{Index terms---}} #1}

\maketitle

\begin{abstract}
An efficient topology management in future 6G networks is one of the fundamental challenges for a dynamic network creation based on location services, whereby each autonomous network entity, i.e., a sub-network, can be created for a specific application scenario. In this paper, we study the performance of a novel topology changes management system in a sample 6G network being dynamically organized in autonomous sub-networks. We propose and analyze an algorithm for intelligent prediction of topology changes and provide a comparative analysis with topology monitoring based approach. To this end, we present an industrially relevant case study on edge video distribution, as it is envisioned to be implemented in line with the 3GPP and ETSI MEC (Multi-access Edge Computing) standards.  For changes prediction, we implement and analyze a novel topology change prediction algorithm, which can automatically optimize, train and, finally, select the best of different machine learning models available, based on the specific scenario under study. For link change scenario, the results show that three selected ML models exhibit high accuracy in detecting changes in link delay and bandwidth using measured throughput and RTT. ANN demonstrates the best performance in identifying cases with no changes, slightly outperforming random forest and XGBoost. For user mobility scenario, XGBoost is more efficient in learning patterns for topology change prediction while delivering much faster results compared to the more computationally demanding deep learning models, such as  LSTM and CNN. In terms of cost efficiency, our ML-based approach  represents a significantly cost-effective alternative to traditional monitoring approaches. 
\end{abstract}

\section{Introduction} \label{sec:Introduction}
The research and development is intensely underway to realize the vision of 6th Generation (6G) wireless networks.  A flexible 6G network is envisioned to encompass the integration of non-terrestrial networks (NTN), device-to-device (D2D) communication, and a vast number of other sub-networks \cite{9625032, flag,10244210}. A generic 3GPP 6G network is composed of core network, access network, and its connected hosts. The core network can be organized in sub-network to provide essential services. The access network connects the hosts to its sub-network. The topological flexibility allows us to incorporate a range of technologies, including 3GPP-compliant radio access networks, Wireless LANs, or even compatible fixed networks.  Finally, hosts that connect to the access sub-network typically consume services provided by the core network. Sub-networks can be defined also based on other criteria, such as on wireless technology, ownership or specific applications.

The dynamic nature of sub-networks creation and maintenance makes the effective topology management increasingly critical. The ability to respond to topology changes is especially essential for maintaining optimal performance and ensuring seamless service delivery. Existing technologies like European Telecommunications Standards Institute (ETSI) MEC \cite{EtsiMec} and Low Latency Low Loss Scalable Throughput (L4S) \cite{IETFL4S}, have addressed the issue of network awareness for applications, recognizing the current limitation in our ability to manage dynamic topology changes, especially in real time. This limitation reduces also the early warning capabilities, thus negatively impacting the system's responsiveness to network fluctuations \cite{10.1145/3409501.3409532}. The sub-network concept is well-aligned with the MEC Zone API. In contrast to past standards based on geo-location information however, the service provisioning in 6G networks is more focused on the UEs that want to use its service and the related topology they are associated with, and less so on the exact location of the UE. The related standardization and research work is still in its infancy. 

In this paper, we propose a novel topology management system for future 6G networks, and illustrate its performance on a practical, industrially relevant case study of edge video distribution. To this end, we first present a video distribution case study, as it is envisioned to be implemented in line with current MEC standards. The novel contribution of our study is to provide a novel topology management system for service oriented 6G architecture, along with a predictive topology detection and changes algorithm, with various ML-based methods that can be dynamically selected based on the service scenarios. We implement the proposed topology management system using an emulation framework, and analyze the impact of topology changes on performance, such as due to variations in sub-network capacity and user equipment (UE) mobility. Based on the key monitoring parameters identified, such as Received Signal Strength Indicator (RSSI), throughput, latency, and mobility, we implement and analyze various ML methods for topology prediction.  Our approach is novel also as it does not only predicts the occurrence of topology changes but also identifies the specific type of change that takes place. 
Additionally, we introduce a cost model to compare traditional monitoring approaches with our ML-based solution, allowing us to assess the cost efficiency of the proposed method.

The remainder of the paper is structured as follows: Section \ref{sec:relwork} reviews the related work. Section \ref{sec:ServiceOriented} presents the reference architecture and the case study. Section \ref{sec:ML-Based} details the algorithm for topology change prediction and the cost model. Section \ref{sec:evluation} presents the performance results. Section \ref{sec:conclusion} concludes the paper.

\begin{figure*}
 \centering
\includegraphics[width=\textwidth]{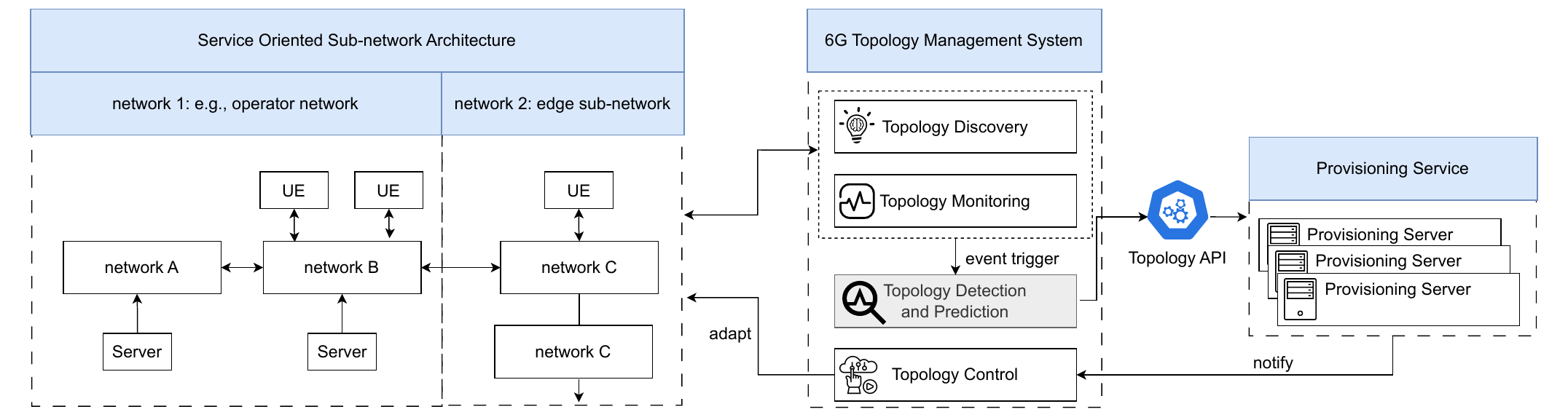}
 \caption{A reference architecture for service-oriented 6G networks.}
 \label{fig:archG}
\vspace{-0.6 cm}
\end{figure*}

\section{Related work} \label{sec:relwork}
A few recent studies, such as \cite{9114878,10038810,9083877}, explored the concept of sub-networks in 6G networks. In \cite{9114878}, the sub-network is presented as a specialized and localized wireless network designed to meet extreme requirements. Additionally, "in-X sub-networks" from paper  \cite{10038810} were engineered to achieve ultra-low latency and extremely high reliability, where 'X' represents the specific environment, or domain (e.g., vehicle, body, or avionics) in which the sub-network operates. Similarly, \cite{9083877} discusses the role of 6G short-range, low-power sub-networks in supporting critical services. Although the concept of sub-networks is gaining attention and some efforts have been made toward standardization, there is still a lack of a common standard, definition, or formal specification from 3GPP. 

The management of sub-network topology was intensely studied in the past in the context of wireless sensor or satellite networks, albeit not from the point of view of the mobile network standard's framework. Some ideas, like in \cite{8253956} are still relevant, however, whereby a wireless sensor network is modeled as a directed graph, using a linear model of noisy measurements. On the other hand, in 6G networks it is expected that in dynamic environments multiple changes might occur over time, unlike in sensor networks studied where typically one change at a time is detected. In our approach, we aim at providing solutions for predicting multiple type of changes rather than just detecting them. We also show that our framework can be used to model the network efficiently, saving time and enabling us to explore multiple change scenarios \emph{a priori}.  Paper \cite{9797135} presents a method for detecting malicious topology changes and localizing affected sensors in wireless sensor networks. The approach relies on data collected at the network's edge and uses a topology distance measure to identify and locate affected sensors. In contrast, our work goes further by incorporating additional parameters, such as RSSI, throughput, latency, and mobility patterns.  Our previous work in \cite{bensalem2024signaling} focused on L3 signaling for mobility in 6G networks, proven to be highly affected by user speed, locations of 6G devices and obstacles, thus strongly motivating the need for topology change detection and prediction. Paper \cite{10659876} present an approach for detecting topology changes in station areas and identifying the type of change, e.g., user migrations into and out of the area, and user phase adjustment. In our scenarios, we adopt a similar approach, by defining: the user remaining in the same area, the user moving outside the area and staying there, and the user moving outside the area and re-entering. Finally, for satellite networks, or network of flying objects, predefined trajectories exist, which simplifies the topology prediction. To the best of our knowledge, no existing studies have addressed the topology change prediction in generic mobile networks using ML tools. 

An increased dynamicity of changes in the network topology, necessarily leads to changes in the application performance. The automation \cite{10244210} to managing the number of network nodes also contributes to this dynamicity. For example, energy management functions modify network capabilities on a regular basis. Consequently, applications might want to either interact with those functions to request capacity or might want to reorganize data flows. The notion of sub-networks that might be moving with vehicles and might be temporarily autonomous, brings dynamic topology changes. Enabling the application layer to react on those makes data flow reorganization possible to maximize QoE \cite{9617565}. A few distinct approaches have been proposed to make the application aware of the changes and events occurring in the network connectivity. For instance,  L4S (Low Latency Low Loss Scalable Throughput) \cite{IETFL4S}  enables applications to gather detailed, real-time information by observing per-packet in-band information about the network performance. There are also efforts to give the application more fine-grained control over how they handle their network connections by extending the current Berkeley socket interface. One such activity is IETF’s TAPS \cite{TAPS}. Additionally, the ETSI MEC \cite{EtsiMec} initiative offers an approach aimed at integrating with the network infrastructure, providing a solution to host application servers closer to the wireless network infrastructure. 

Relevant to topology management of sub-networks is ETSI MEC standard that provides a set of APIs for key MEC interfaces to query network status, and in particular Location API (LAPI) \cite{LocAPI}, Traffic Management APIs \cite{BWAPI}, and Radio Network Information API \cite{RadioAPI}. A salient feature of this LAPI, which is different from MEC Zone API, is that the service provisioning service, in the context of topology management is more focused on the UEs that want to use its service, and less so on the exact geo-location of the UE, as it is traditionally the case. The ETSI MEC framework also enables mobile wireless clients that may transition between compute resources, as the compute resources serving the application clients may not be accessible from all client communication domains. In this context, the topology management that encompasses compute and communication resources as well as enables integration and compatibility with ETSI MEC standard is the central goal of the approach shown in this paper.

\section{Service Oriented 6G Networks}\label{sec:ServiceOriented}

\subsection{Reference Architecture}
Fig. \ref{fig:archG} illustrates  the reference architecture for service oriented sub-networks, where we emphasize 6G topology management system, Topology API (TAPI) and service provisioning modules in the network management planes. Fig. \ref{fig:archG} illustrates one edge and one operator sub-network, with their corresponding user equipment, servers and hosts.
\par The 6G Topology Management System is envisioned to discover, monitor, detect, predict and control the subnets, enabling close to real-time adaptation to changing network conditions in various sub-networks. 
This paper focuses particularly on two distinct functions: Topology Monitoring, which operates as a reactive, monitoring-based approach to detect topology changes through real-time measurements of network parameters, and Topology Detection and Prediction, which uses an ML-based approach to leverage information about network changes, such as fluctuations in sub-network capacity, significant variations in connectivity characteristics, and user or sub-network mobility, to proactively predict topology changes, as described in Section \ref{sec:ML-Based}.

\par We propose the  Topology API between the topology management system and the service provisioning. It provides close-to-real-time information about the topology changes. This API enables the provisioning service to make informed decisions about the dynamic association of entities within the system (e.g., UE to servers association). TAPI is closely related to the MEC Location API and Zone API standard \cite{LocAPI}. 
We emphasize however that TAPI is different from MEC Zone API, as it is more focused on the UEs that want to use its service and need to be topologically embedded in the sub-network, and less so on the exact geo-location of the UE. 

\par Finally, the provisioning service management oversees the association of users to sub-networks, dynamically updating them when changes in topology, network capacity, or user movement are detected. It notifies the sub-network associations through the Topology Control  functions from the 6G Topology Management System. It is implemented in distributed (albeit logically centralized) provisioning servers (PS).
\begin{figure}
\centering
\includegraphics[scale=0.45]{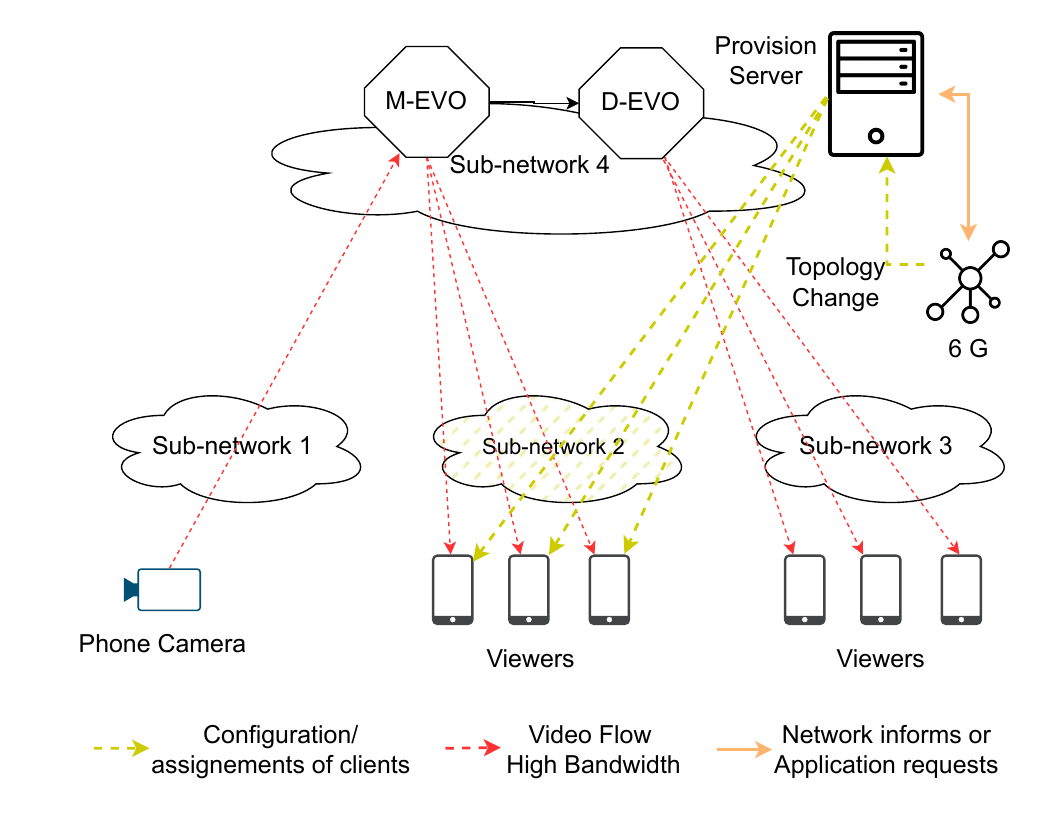}. 
\caption{Edge video distribution application in 6G sub-networks.}
\label{fig:usecase}
\end{figure}
\subsection{Case Study: An Edge Video Distribution}
\label{casestudy}


\begin{figure*}
    \centering
    \begin{subfigure}[b]{0.45\linewidth}
        \centering
        \includegraphics[width=\linewidth]{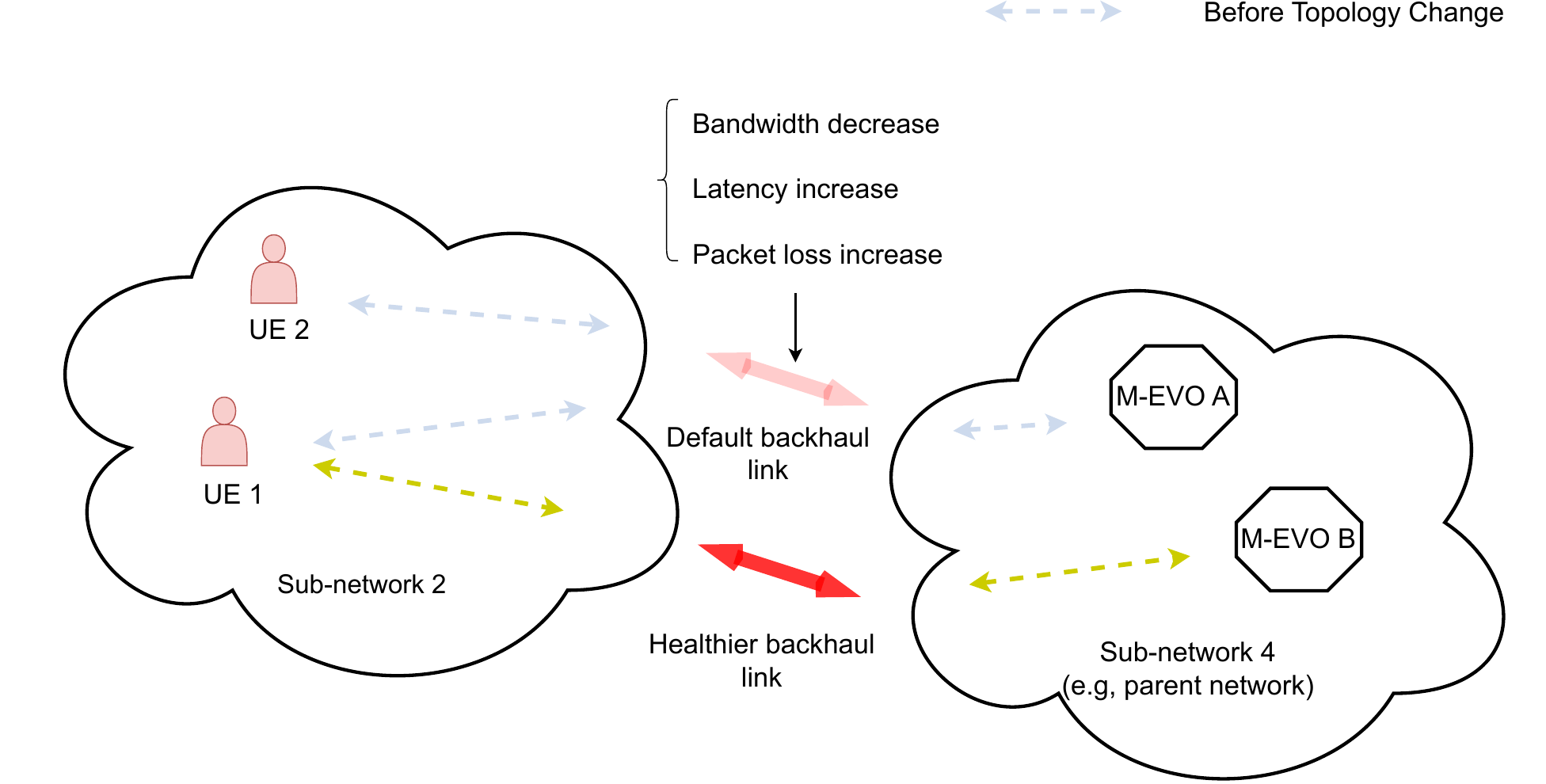}  
        \caption{Scenario A: sub-network's parent link change considerably. }
        \label{fig:subfig-a}
    \end{subfigure}
    \begin{subfigure}[b]{0.45\linewidth}
        \centering
        \includegraphics[width=\linewidth]{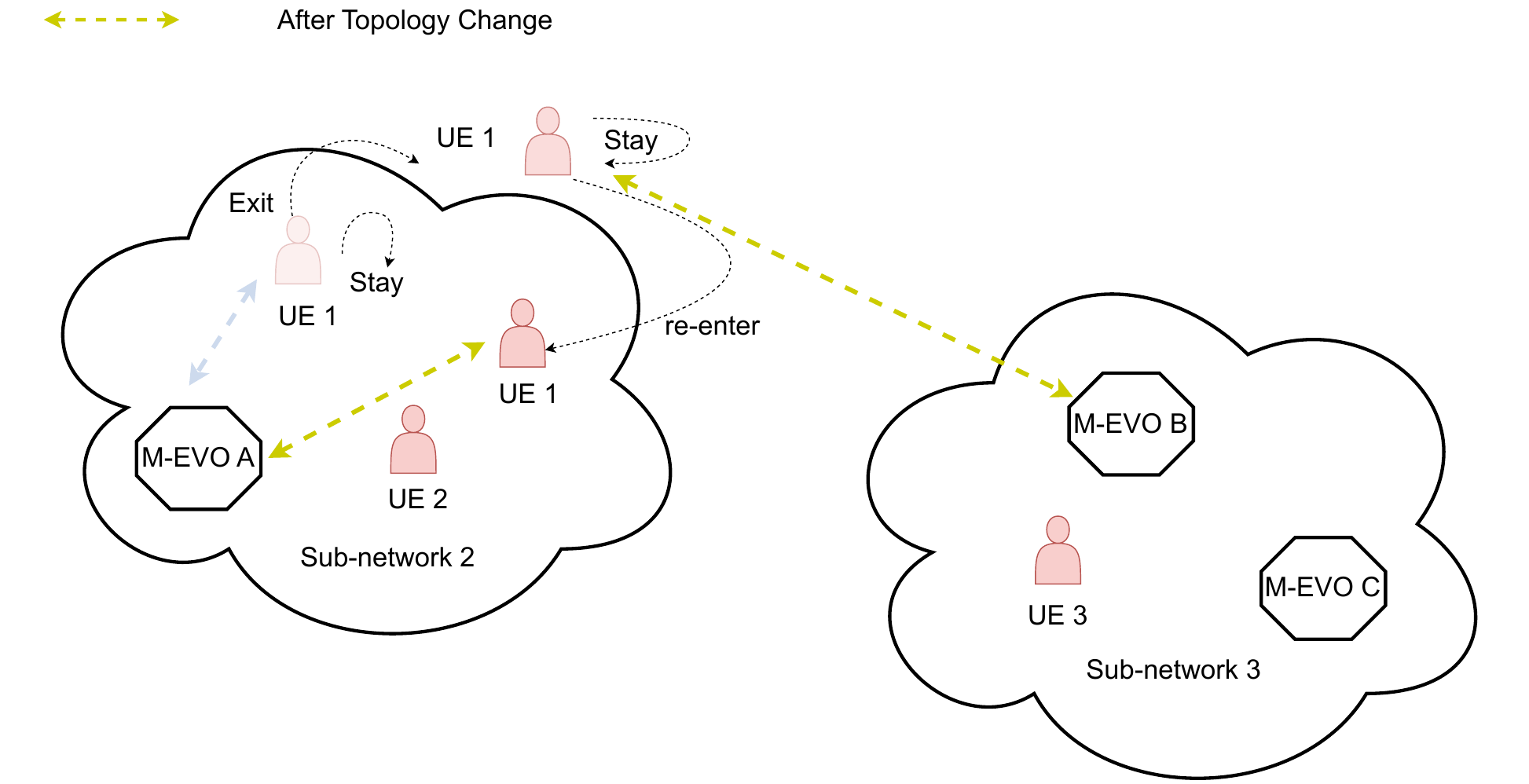}  
        \caption{Scenario B: UE stay, exit or re-integrate.}
        \label{fig:subfig-b}
    \end{subfigure}
    \caption{Topology changes exemplified.}
    \label{fig:TopologyChangesExamplified}
\end{figure*}

We describe a real-world, industrially relevant case study of edge video distribution, shown in Fig. \ref{fig:usecase}. Let us assume a video streaming application along with telemetry data stream distribution, whereby several parallel streams are generated in one area of the network (stream providers) and then distributed to a high number of users consuming these streams (stream receivers). The stream receivers can be in several remote sub-networks; whereas Phone Camera is in Sub-network 1, the viewers are in Sub-networks 2 and 3. Sub-network 4 acts as a \emph{parent network}, in which it provides connectivity between all other sub-networks. For streaming, the recorded streams are first sent to a server (Master Edge Video Orchestrator (EVO), M-EVO). This server then forwards the streams to one or more distribution servers (Distributor EVO, D-EVO). It is  the task of D-EVOs to replicate the streams to each client device. This management of individual streams is orchestrated by a PS.

This paper examines two scenarios, as shown in Fig.\ref{fig:TopologyChangesExamplified}, i.e.,
\subsubsection{Scenario A} The parent network connectivity undergoes a significant performance change, e.g., via the backhaul (sub-network 4) with severely reduced bandwidth, increased latency, or packet loss. This triggers a re-association between UEs (i.e., viewers
) and EVOs (be it M-EVO or D-EVO). As shown in Figure \ref{fig:TopologyChangesExamplified}a, when UE 1 and UE 2 in sub-network 2 are connected to M-EVO A via a degraded link, this degradation prompts a re-association. The re-association may lead to UEs connecting to a backhaul link with better quality, resulting in a new association between UE 1, UE 2, and M-EVO B. 

\subsubsection{Scenario B}  The UE moves between different areas, either defined as another sub-network or as an area without reachability. UE 1 moves between sub-network 2 and an external area. Sub-network 2 contains an EVO and several UEs, with UE 1 either staying in sub-network 2, or exiting to connect to another sub-network (sub-network 3). The Exit arrow indicates UE 1 moving out of sub-network 2, which could trigger a disconnection from its current EVO. In contrast, the Stay arrow represents UE 1 remaining connected to M-EVO A in sub-network 2. The exit of UE 1 from sub-network 2 presents a topology change, requiring a re-association between UEs and EVOs (i.e., UE 1 connecting to M-EVO B). The re-entering path illustrates a scenario where UE 1 leaves sub-network 2 but later re-enters and reconnects. In this case,  UE's mobility triggers another topology change, once again affecting its association with the local EVO (e.g., re-association with M-EVO A).


\section{Topology Change Monitoring and Prediction} \label{sec:ML-Based}
In this section, we focus on the  functioning of the Topology Monitoring, Detection and Prediction modules proposed in Fig. \ref{fig:archG}. It should be noted that the topology changes prediction is not a necessary module and the topology changes can be detected also via simple monitoring. We propose to this end a simple cost model to compare both a traditional monitoring based approach and our predictive, ML-based approach, in order to evaluate the effectiveness of both solutions.

\subsection{Monitoring Based Approach}
Topology monitoring is a fundamental process for managing dynamic network environments, particularly in 6G networks characterized by autonomous sub-networks. It involves continuously tracking network parameters, such as link quality, bandwidth, and delay, to identify and respond to topology changes in real time. Unlike topology change prediction, which uses advanced algorithms to anticipate changes, topology monitoring relies on direct measurements and observations to detect changes as they occur.
\subsection{Topology Detection and Prediction}\label{MLbasedChangeDetectionandPredictionSolution}

\begin{figure*} 
 \centering
   \includegraphics[scale=0.3]{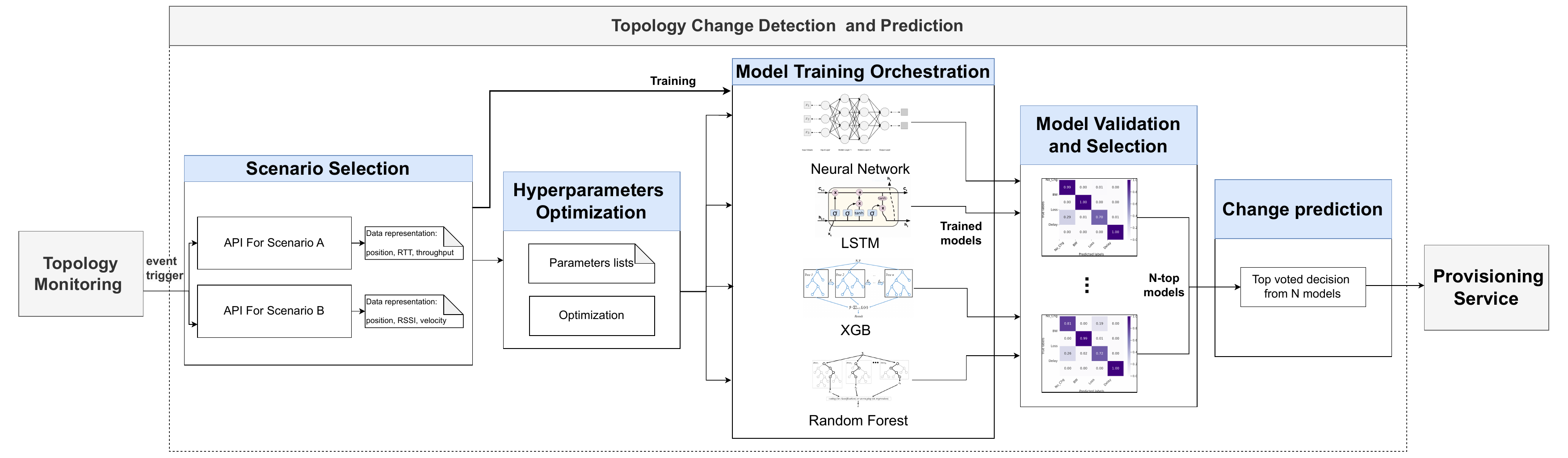}
 \caption{ ML-based change detection architecture.}
\label{fig:arch_AI}
\end{figure*}

We exploit several ML-based approaches and different parameterized Deep Neural Networks (DNN) architectures to implement topology change detection and prediction of future changes for the case study proposed. This is illustrated in Fig. \ref{fig:arch_AI}. The network monitoring component triggers the network status updates, such as throughput, Round Trip Time (RTT) and RSSI. The module called \emph{Scenario Selection} receives monitoring data collected from the network, and identifies the parameters required (Scenario A and B).  For each scenario, we implement an API to be triggered internally to filter the received data from the monitoring event, check if there is any modification to the needed parameters, and then extract the attributes needed for the topology change detection.

Before training ML models and more specifically neural network based models, it is essential to use the best fitting hyperparamters for the problem studied (\emph{Hyperparameter Optimization}). The related component generates a  list of possible combinations of model hyperparameters, and evaluates several combinations using some data samples to find the most efficient combination of hyperparamters. Then, the meta-data is sent to the \textit{Model Training Orchestration} for training. 

This component receives the data and the corresponding scenario from \textit{Scenario Selection}, and the hyperparameters for each ML model from the \textit{Hyperparameter Optimization}. The selected ML models to be used for all scenarios are: {Artificial Neural Network (ANN)}, which is the basic fully connected neural network, { Convolutional Neural Network (CNN)} are a class of deep learning models primarily used for analyzing visual data, though they can also be applied to other types of structured data, The {Long-Short term memory (LSTM)} Neural Network, which was  built to capture temporal patterns and performs exceptionally well in time-series forecasting and will be used mainly to predict future topology changes in scenario B, { Extreme Gradient Boosting (XGB)} is an optimized, high-performance implementation of the gradient boosting algorithm. It is designed for speed and efficiency, often outperforming other ML models on structured data, and finally {Random Forest (RF)} is an ensemble machine learning algorithm that operates by constructing multiple decision trees during training and combining their results to improve accuracy and prevent overfitting. After training all models, the model files are saved for testing and exploitation.


After the training, all models are evaluated with \textit{Testing Data}, which is a labeled data collected most recently in the measurement. The MSE and accuracy are calculated and then the N-top performing models are selected, and assigned to predict changes for each scenario.

 The algorithm with Top-N model voting is presented in \textbf{Alg.} \ref{alg:CPvoting}. The selected models are used to predict changes every time the module is triggered with new events and the decision that is most voted by all models is decided with a priority to the model with highest accuracy and least MSE. In order to save time and ensure accuracy, only the Top-N models are required to evaluate the data and conclude the change decision.   
\begin{algorithm}
\caption{Change Prediction with Top-N Model Voting}\label{alg:CPvoting}
\begin{algorithmic}[1]
\Require $\text{TopNModels}$: A list of Top N models
\Require $\text{Input}$: Input data for decision
\Ensure Final decision based on majority vote

\State \textbf{Initialize} $\text{Votes} \gets \{\}$ \Comment{Empty dictionary to store votes}

\For{$i = 1$ to $N$}
    \State $\text{Decision}_i \gets \text{TopNModels}[i].\text{predict(Input)}$
    \If {$\text{Decision}_i \text{ in Votes}$}
        \State $\text{Votes}[\text{Decision}_i] \gets \text{Votes}[\text{Decision}_i] + 1$
    \Else
        \State $\text{Votes}[\text{Decision}_i] \gets 1$
    \EndIf
\EndFor

\State \textbf{Select} the decision with the highest count in $\text{Votes}$ as $\text{FinalDecision}$
\State \Return $\text{FinalDecision}$
\end{algorithmic}
\end{algorithm}
\subsection{ML vs Monitoring based approach}
There are cost implications of utilizing an ML approach versus traditional network monitoring tools to detect and predict changes within a network. The ML approach described previously and can be deployed as a containerized application in the network, while monitoring approach is solution dependent which might need dedicated hardware, or servers to run network monitoring services.  We now evaluate both approaches in terms of total cost that covers different factors such as  purchase costs, annual maintenance fees, operational costs, usage levels, subscription fees, scope of deployment, monitoring requirements, network size and energy consumption, which are used as examples that can help determine which approach is the most cost-effective.

Network monitoring tools come in  different forms to track bandwidth, latency, and packet loss, each suited to specific needs. Three types of solutions are defined: High-end hardware,  Software-based, and Cloud-based. 

\emph{High-end hardware solution}. We denote by $N_{\text{device}}$ the number of hardware devices deployed at each node, $C_{\text{device}}$ the cost of each device, and $C_{\text{maintenance}}^{\text{HW}}$ the annual maintenance cost per device. The hardware-based solution cost is defined as follows:
\begin{equation}\begin{split}
    Cost(\text{HW}) = & N_{\text{device}} (C_{\text{device}} + C_{\text{maintenance}}^{\text{HW}} )
\end{split}
\end{equation}
\emph{Software-based solution}. We denote by  $C_{\text{setup}}$ the cost of setup of the software solution, $C_{\text{subscription}}$ the subscription per year,  $N_{\text{VM}}$ the number of virtual machine (VM) used for deployment, $C_{\text{VM}}$ the cost of VM allocation, and $C_{\text{maintenance}}^{\text{SW}}$ the annual maintenance cost of the software (SW). The software-based solution cost is defined as follows:
\begin{equation}\begin{split}
    Cost(\text{SW})  = & C_{\text{setup}} + C_{\text{subscription}} + N_{\text{VM}} C_{\text{VM}} + C_{\text{maintenance}}^{\text{SW}}
\end{split}
\end{equation}
\emph{Cloud-based solution}.  We denote by $C_{\text{subscription}}$ the subscription cost per year, $N_{\text{end-point}}$ the number of end points being monitored, $C_{\text{end-point}}$ additional cost per monitored end-point, and $C_{\text{storage}}$ cost of storing data. The cloud-based solution cost is defined as follows:
\begin{equation}\begin{split}
    Cost(\text{cloud})  = &  C_{\text{subscription}} + N_{\text{end-point}} C_{\text{end-point}} + C_{\text{storage}}
\end{split}
\end{equation}
\textbf{Monitoring-based Approach} 
The cost of this approach  is composed of the sum of two parts:  Capital Expenditure (CapEx) which refers to fixed costs, Operational Expenditure (OpEx) which refers to the ongoing expenses for running the system, mainly here energy consumption. The first part defined by $C_{M}(\text{type})$ is a cost function that incorporates a parameter type, which represents one of three monitoring solutions provided previously, which are reflected in the cost function based on the selected parameter. The second part is defined by $E_{total}(N_{\text{server}}, \lambda_{\text{mon}}, \mu_{\text{mon}})$.  The cost of this approach is:
\begin{equation}\begin{split}
    C_{M}(\text{type})= &  Cost(\text{type})
     +   E_{total}(N_{\text{server}}, \lambda_{\text{mon}}, \mu_{\text{mon}})
\end{split}
\end{equation}
Except for the cloud-based monitoring, the energy consumption is not considered as we assume that the cloud is not owned by the network.\\
\textbf{ML-based Approach} The cost of this approach is mainly consisting of three parts: the maintenance costs, the training costs and the inference costs. As the ML models can be containerized and deployed in any existing server in the network, the maintenance costs will include the cost of deploying and maintaining the pods, while  both training and inference costs will  consider the energy consumption of each process. 
\begin{equation}\begin{split}
    C_{ML}= &   C_{\text{maintenance}}^{\text{ML}} + E_{total}(N_{\text{train\_server}}, \lambda_{\text{train}}, \mu_{\text{train}}) \\    & + E_{total}(N_{\text{inference\_server}}, \lambda_{\text{inference}}, \mu_{\text{inference}})
\end{split}
\end{equation}

To calculate $E_{total}$, we  denote by $P_{\text{idle}}$ and $P_{\text{peak}}$ the idle and peak power of a server, respectively,  $E_{\text{usage}}$ the power consumption for cooling, and other overhead of turning on a server, $N_{\text{server}}$ the number of running servers. We adopt an $M/M/1$ queuing system to analyze the behavior of a server and to compute the CPU utilization $U(\lambda, \mu)= \frac{\lambda }{ N_{\text{server}} \mu}$ according to the workload, where $\lambda$ and $\mu$ denotes the arrival and service rates of the workload, respectively. Similar to  \cite{kiani2016profit}, the total energy consumption is calculated as follows:
\begin{equation}\begin{split}
    E_{total}(N_{\text{server}}, \lambda, \mu)= & N_{\text{server}} \left[ \left( P_{\text{idle}} + [E_{\text{usage}} - 1]P_{\text{peak}} \right) \right.\\ 
   & \left.+ (P_{\text{peak}} - P_{\text{idle}})U(\lambda, \mu) \right]\\
\end{split}
\end{equation}


\section{Performance Evaluation}\label{sec:evluation}

\subsection{Evaluation setup}\label{subsec:evluationsetup}
The primary challenge in the evaluation setup is to create a testing environment that closely emulates our  case study, ensuring that the results obtained, particularly regarding throughput, RTT and RSSI are meaningful and applicable. To this end,  we configure Mininet-wifi \cite{7367387} to implement the concept of sub-networks, parent network, UE and EVOs. We apply the following configurations to the case study analyzed. 

\begin{figure}
 \centering
  \includegraphics[scale=0.6]{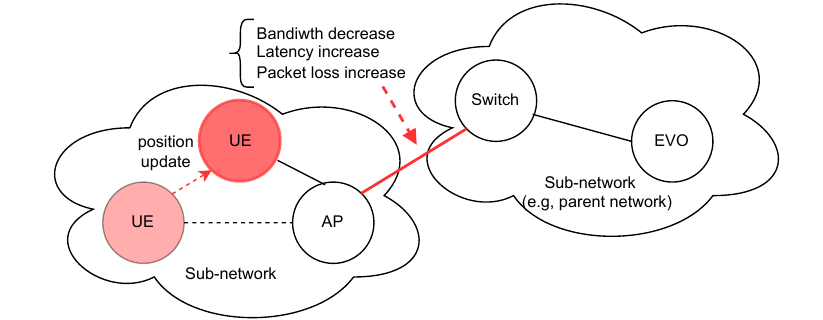}
 \caption{Topology emulated in Scenario A.}
\label{fig:TopoSc1}
\end{figure}

In Scenario A, we assume that the sub-network is connected to the parent network using an access network-to-core network link. We furthermore assume a simple parent network that includes only core network functionality, which is linked to a sub-network containing hosts and access network functionality. Here, there is no network autonomy, i.e., the sub-network relies on a connection to the parent network via a backhaul link for full functionality. As a result, we configure the topology illustrated in Fig. \ref{fig:TopoSc1}, where we see that a sub-network is connected to another sub-network (i.e, parent network) via a backhaul link between the access point and the switch (SW). In the default state, the link operates without any restrictions, allowing for best performance. To simulate degradation in the link, we adjust the bandwidth, delay, and packet loss to create scenarios that impact network performance. We use \texttt{iperf} measure throughput and \texttt{ping} to assess latency between a UE client (station) in one sub-network and an EVO server (host) in the parent network. We assume that the client is moving, dynamically changing its location. The relevant simulation parameters for this scenario are detailed in Table \ref{tab:Sc1Par}.
\begin{table}[htb]
\centering
\caption{Emulation parameters in Scenario A.}
\begin{NiceTabular}{|ll|c|}
\hline
 \multicolumn{2}{c}{\textbf{Parameter}}              & \textbf{Value}                             \\ \hline
Bottlenecks       &     Bandwidth               & 0.1Mbps            \\ 
           & Packet Loss               & 60\%        \\ 
           & Delay                         &  100ms       \\ \hline
Position change      &                  &  1.0 m/s  \\ \hline
Access Point &          channel         & 1  \\
             & mode & g                                    \\ \hline
Number of iterations        &     & 5k                                        \\ \hline
\end{NiceTabular}%

\label{tab:Sc1Par}
\end{table}

\begin{figure}
 \centering
  \includegraphics[scale=0.6]{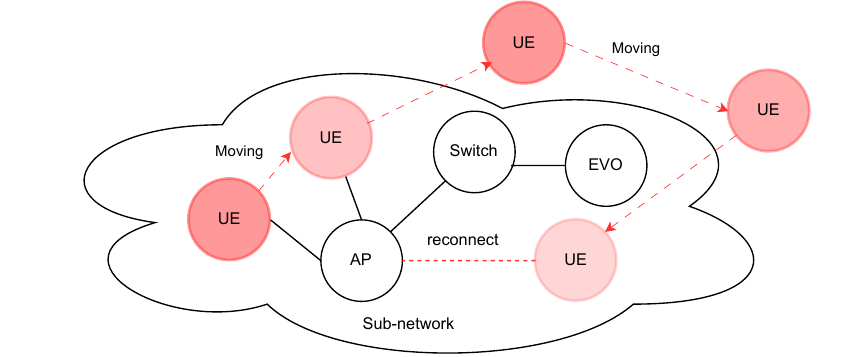}
 \caption{Topology emulated in Scenario B.}
\label{fig:TopoSc2}
\end{figure}

In Scenario B, we assume a fully autonomous sub-network with autonomous network management functions, including hosts, access network, and core network components, i.e., without any reliance on a connection to the parent network. The emulated topology is shown in Fig. \ref{fig:TopoSc2}, where we emulate an UE remaining, exiting, or re-entering into a sub-network by utilizing the mobility features provided by the emulation tool. The concept of a sub-network in this context is represented by the coverage area of an access point. By moving a station (representing the UE) to different positions, we simulate the UE transitioning between being inside and outside the sub-network. Since movement outside the AP's coverage area constitutes a significant topological change, we define the coverage area itself as a reference point, which serves as an early indicator of such transitions.

In this study, we assume that both the UE client and the EVO server reside within the same sub-network. We model a mobile network where a station (node, or UE client) moves randomly within a defined area using an updated version of the Random Walk mobility model provided by the Mininet emulation framework. The key parameters used in our simulation are listed in Table \ref{tab:Sc2Par}. After each movement, the station’s position, distance to the access point, RSSI, velocity, and coverage status are updated and logged for later analysis. The system monitors the position and signal strength of a station to determine if it is inside or outside the AP coverage area.

\begin{table}[htb]
\centering
\caption{Emulation parameters in Scenario B.}
\begin{NiceTabular}{|ll|c|}
\hline
 \multicolumn{2}{c}{\textbf{Parameter}}              & \textbf{Value}                             \\ \hline
Propagation      &   model               & log-distance propagation        \\  
           & path loss exponent                         &  5       \\ \hline
Velocity      &                  &  1.0 - 5.0 m/s  \\ \hline
Mobility model & direction  (x, y)    & uniform(-1, 1)                             \\ \hline
Access Point &      channel         & 1   \\ 
             & mode & g                                    \\ \hline
Number of iterations        &     & 10k                                        \\ \hline
Noise Threshold        &     & -91                                        \\ \hline
Fading Coefficient        &     & 0                                        \\ \hline

\end{NiceTabular}%

\label{tab:Sc2Par}
\end{table}

\subsection{Performance Results}
We now presents the performance results obtained and analyze the efficiency of the proposed methods by also comparing it to monitoring based methods.

\begin{table} \small%
\caption{ML model settings.}

\begin{NiceTabular}{|p{0.04\textwidth}|p{0.12\textwidth}|p{0.04\textwidth}|p{0.18\textwidth}|}
\hline
\multirow{1}{*}{\textbf{ML }} & \multicolumn{1}{c}{\textbf{Scenario A}} & \multirow{1}{*}{\textbf{ML }} & \multicolumn{1}{c}{\textbf{Scenario B}} \\
   \multirow{1}{*}{\textbf{Model }} & \multicolumn{1}{c}{Hyperparameters} &  \multirow{1}{*}{\textbf{Model }}& \multicolumn{1}{c}{Hyperparameters}  \\
\hline
ANN         &   solver='lbfgs',                & CNN & solver='adam' \\
            &   alpha=1e-5,                    &     & filters=64, kernel\_size=1, \\
            &   hidden\_layer\_                &     & Flatten,  LeakyReLU,  \\
            &    sizes =(3, 8, 8)              &     & BatchNormalization,   \\
            &                                  &     & Dense=1, batch\_size=4   \\
\hline
RF          &  n\_estimators=40,               &LSTM & solver='adam', \\
            &  max\_depth=7                    &     & units=64, LeakyReLU, \\
            &                                  &     & BatchNormalization,  \\
            &                                  &     & Dense=1, batch\_size=4 \\
\hline 
XGB         &  n\_estimators=40,               & XGB & n\_estimators=20, \\
            &  max\_depth=6                    &     & max\_depth=2\\
\hline 
\end{NiceTabular}
\label{tab:modelconf}
\end{table}

\begin{figure*}[ht]
 \centering 
 \begin{subfigure}[t]{0.32\linewidth}
 \includegraphics[scale=0.34]{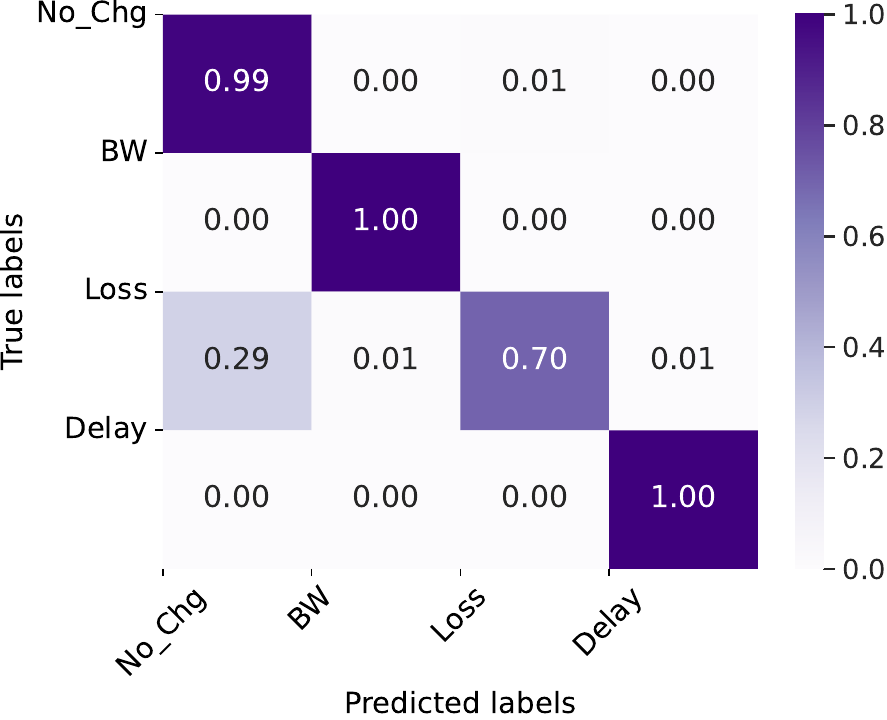}
  \caption{ANN  }\label{subfig:mat1}
   \end{subfigure}
   \begin{subfigure}[t]{0.32\linewidth}
 \includegraphics[scale=0.34]{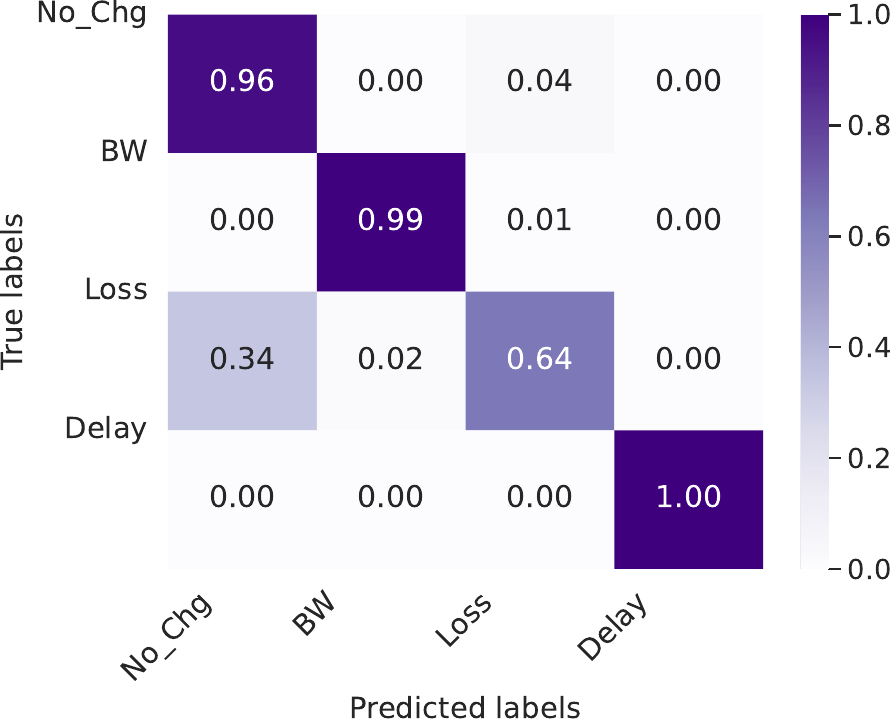}
 \caption{Random Forest}\label{subfig:mat2}
   \end{subfigure}
 \begin{subfigure}[t]{0.32\linewidth}
 \includegraphics[scale=0.34]{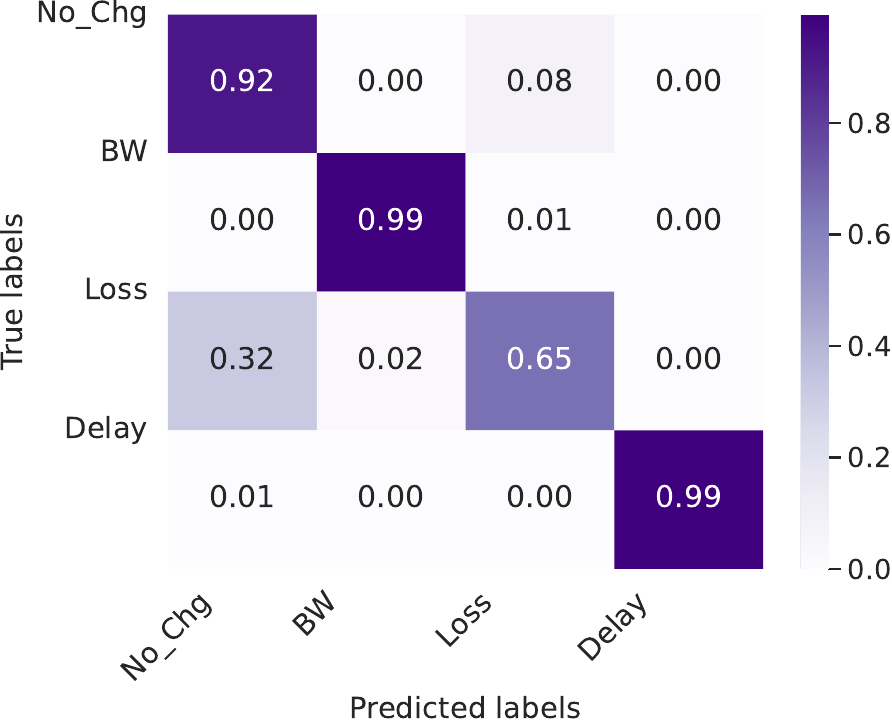}
 \caption{XGB  }\label{subfig:mat3}
   \end{subfigure} 
 \caption{Confusion matrix for NN, Random Forest and XGB.}
\label{fig:confmat}
\end{figure*}

The data used for our ML algorithms is collected from the emulation described in Section \ref{subsec:evluationsetup}. We adopt five ML/DL models for implementation, as already mentioned in Section \ref{MLbasedChangeDetectionandPredictionSolution}. After optimizing the hyperparameters with the   \textit{Hyperparameter Optimization} component, we obtain the configurations described in Table \ref{tab:modelconf}. 
To evaluate our selected set of ML/DL based models, we adopt the most common metrics to validate the detection and classification performance of the trained models, which are the accuracy and the mean square error (MSE). To evaluate the time performance of models, we calculate the training and inference time for each ML model.

For Scenario A, the data collected from the emulation for training is representing the UE position coordinates, the average throughput in Mbit/sec and the average RTT in milliseconds (ms). Each data point is labelled automatically for training based on the change status of the link. We set a constant value considered as a non-change denoted as 0, a change in the bandwidth as 1, in the packet loss as 2, and in the link delay as 3. The training and testing evaluation  performance are presented in Table \ref{tab:algo}. The results shows that ANN is the most accurate model with 92.8 \%, while RF and XGB performance are slightly less accurate. The drawback of ANN appears in the high training time 2.15 s compared to RF which is the lowest 0.15 s. For inference time ANN is also the fastest model per request with 0.24 $\mu$s per data point, meaning that ANN can make 5 million evaluation per second, while the inference for RF is 7.3 $\mu$s. 
Furthermore, as we identify the specific type of change that takes place using our approach,  we evaluate the  confusion matrix for each ML model to evaluate the performance of each model at detecting the existence of a change, the absence of change in the topology is denoted as  \textit{No\_Chg}, and  each type of change: bandwidth change (BW),  packet loss change (LOSS), and link delay change (Delay), as shown in Fig. \ref{fig:confmat}. The three models detect the change in the link delay and the bandwidth  with an accuracy $\geq 99\%$. We observe that ANN outperforms other models for detecting that no change happened in the link with 99\%, while RF and XGB has a slightly lower accuracy 96\% and 92\%, respectively, which confuses it with the status when there is a LOSS change. The accuracy of LOSS detection is less than 70\% for all the models, which shows that it is harder to differentiate between a status with LOSS change and with no-change. 
In conclusion, since maintaining optimal performance and reliability on the backhaul link is crucial, ANN’s advantages in accuracy, MSE, and inference time make it the preferred choice.

For Scenario B, the gathered data  from the emulation is representing the UE position coordinates, the RSSI in dBm, and the velocity of the UE in meter per second (m/s). Each data point is labelled automatically for training based on the change status of the UE. The status is represented by four states. $Change=0$ indicates the default state where the station is inside the sub-network with no significant status change. $Change=1$ occurs when the station exits the coverage area, signaling that it has moved from inside to outside the sub-network. $Change=2$ represents the station remaining outside the coverage area, with no transition back inside. Finally, $Change=3$ indicates that the station has re-entered the coverage area, moving back inside the sub-network after being outside. The goal of ML models is to predict the future UE status using the history of movement and RSSI values.  The training and testing evaluation  performance are illustrated  in Table \ref{tab:algo}. The results shows that XGB slightly outperforms LSTM and CNN in terms of accuracy 95.4\% with a very high advantage in terms of training and inference speed, with inference time  0.9   $\mu$s per data point,  while the inference for LSTM is 175.2 $\mu$s, and 78.1 $\mu$s for CNN. We conclude that XGB as a simple ML model could learn the patterns from the change prediction problem with high accuracy and provides a very fast inference time  compared to DNN based models, which builds complex neuron structures.
Since XGB outperforms other models across all evaluation metrics, it is the preferred choice for user mobility prediction.


\begin{table} \small
\centering
\caption{Algorithms performance.}
\label{tab:algo}

\begin{tabular}{|c|c|c|c|c|}
\hline
\textbf{Scenario} & \textbf{Metrics} & \textbf{ANN} & \textbf{RF} & \textbf{XGB} \\
\hline
A & \textbf{Accuracy} & \textbf{0.928} & 0.907 & 0.895 \\
\cline{2-5} 
  & \textbf{MSE} & \textbf{0.035} & 0.045 & 0.048 \\
\cline{2-5} 
  & \textbf{Training time [s]} & 2.154 & \textbf{0.156} & 0.279 \\
\cline{2-5} 
  & \textbf{Inference time [$\mu$s]} & \textbf{0.24} & 7.3 & 2.8 \\
\hline
\rowcolor{gray!25} B & \textbf{Metrics} & \textbf{LSTM} & \textbf{CNN} & \textbf{XGB} \\
\hline
  & \textbf{Accuracy} & 0.942 & 0.922 & \textbf{0.954} \\
\cline{2-5} 
  & \textbf{MSE} & 0.123 & 0.124 & \textbf{0.112} \\
\cline{2-5} 
  & \textbf{Training time [s]} & 28.53 & 21.26 & \textbf{0.41} \\
\cline{2-5} 
  & \textbf{Inference time [$\mu$s]} & 175.2 & 78.1 & \textbf{0.9} \\
\hline
\end{tabular}

\end{table}


We finally analyze the efficiency of our proposed methods by a samle cost analysis and comparison with monitoring based approach. For each turned-on server, we assume that   $P_{\text{idle}}=0.2$ kw,  $P_{\text{peak}}=0.1$ kw,   and $E_{\text{usage}}=1.2$.
Fig. \ref{fig:confmat} provide a cost evaluation comparison between the monitoring approaches and the ML approach.
Fig. \ref{subfig:mat1} shows the cost comparison among three different types of  monitoring solutions. The hardware-based solution has the highest initial cost and increases as the number of monitored elements grows, which reflects the scaling cost associated with deploying additional hardware. The software solution has a high upfront cost, but its cost grows at a slow, linear rate as the number of monitored elements increases. This indicates that this solution has a high setup cost and adding more elements has a minimal impact on the overall cost. The cloud-based solution starts with the lowest initial cost and  remains low as more elements are monitored. This reflects that the cloud solutions are subscription-based cost model, which is designed to accommodate scaling needs with minimal additional costs per monitored element.

We assume that one server is dedicated to training, with the service rates for training and inference of each ML approach provided in Table \ref{tab:algo}. Fig. \ref{subfig:mat2} illustrates the cost associated with the ML approaches based on the number of pods used for inference. ANN has the lowest cost among the ML approaches and scales linearly with the number of pods. XGB shows a slightly higher initial cost than ANN but scales in a similar linear fashion. RF shows the highest cost increase among the ML solutions. To summarize, ML-based solutions could offer a cost-efficient alternative to traditional monitoring tools, especially as the number of monitored elements increase.


\begin{figure}
    \centering
    \begin{subfigure}[b]{\columnwidth}
        \centering
        \includegraphics[scale=0.27]{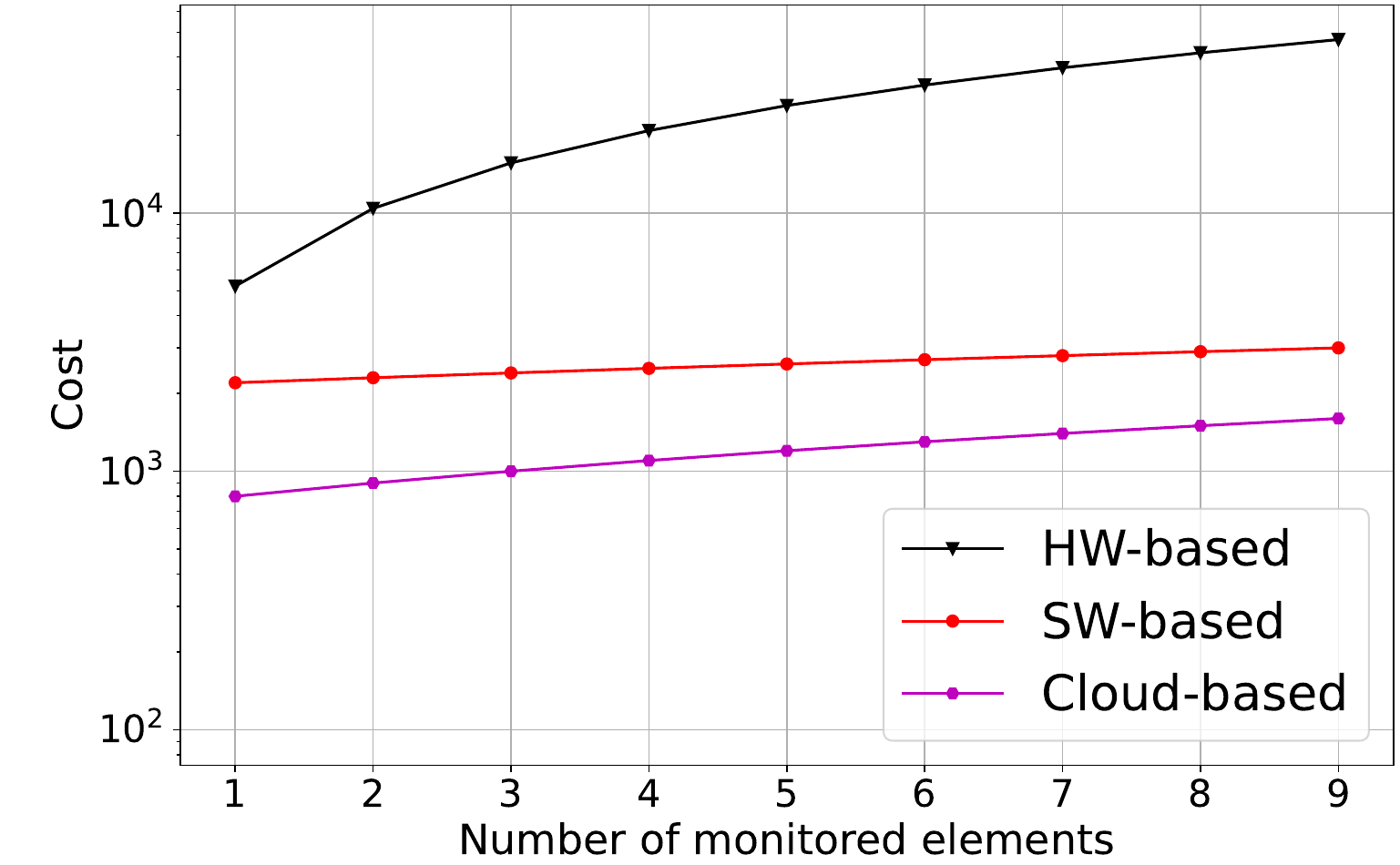} 
        \vspace{-0.1cm}
        \caption{Monitoring approaches.}
        \label{subfig:mat1}
    \end{subfigure}
    \begin{subfigure}[b]{\columnwidth}
        \centering
        \vspace{0.1cm}
        \includegraphics[scale=0.27]{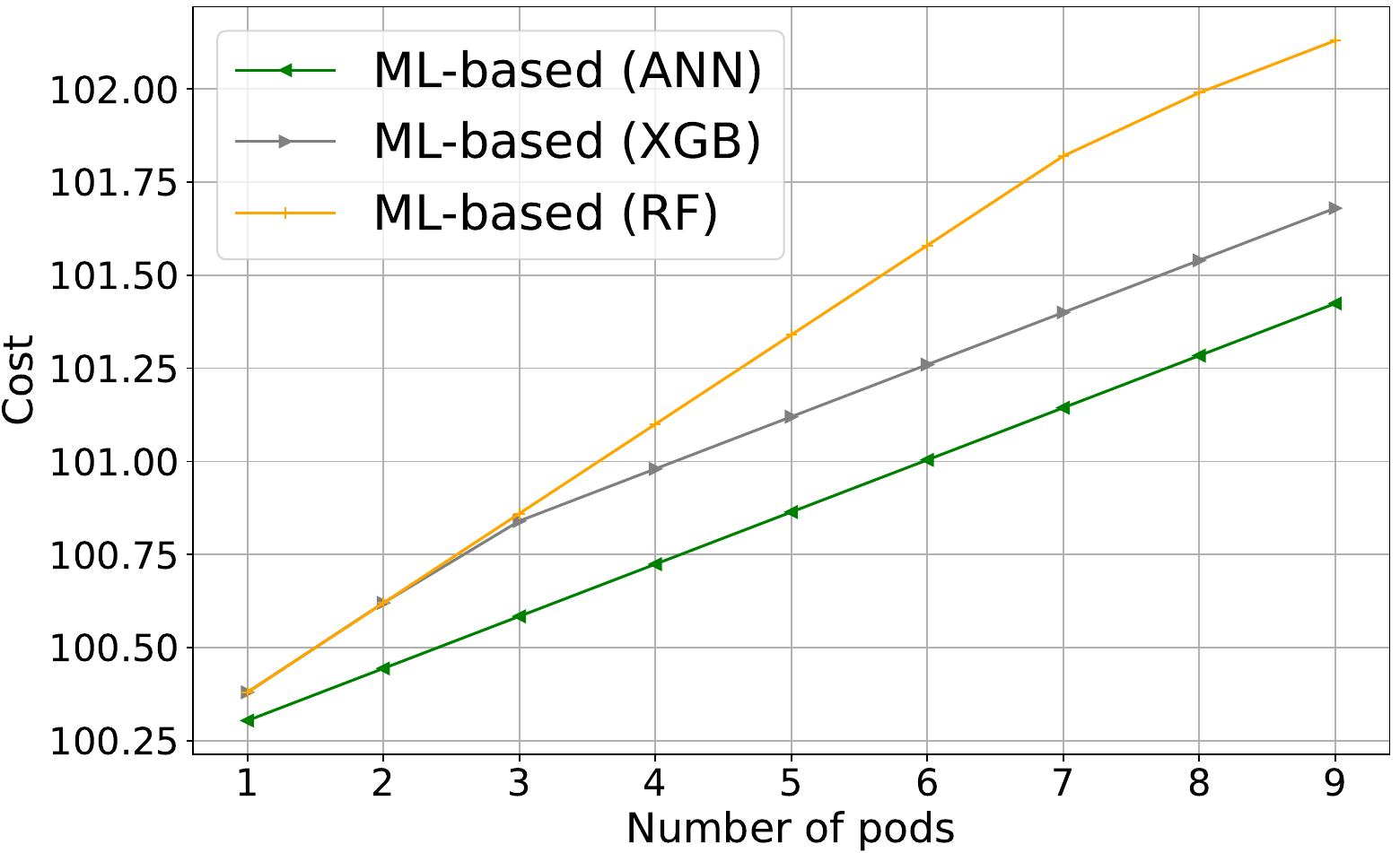} 
        \caption{ML approaches.}
        \label{subfig:mat2}
    \end{subfigure} 
    \caption{Cost efficiency analysis.}
    \label{fig:confmat}
\end{figure}

\subsection{Discussion}
The applicability of our results is subject to the below factors:
\subsubsection{TCP} 
We chose to use TCP only for video distribution to focus on topology change detection rather than detailed performance metrics, avoiding UDP even though it's common for video streaming. 
\subsubsection{Backhaul Link} 
We recognize that the backhaul link is typically characterized by high bandwidth, low latency, and minimal packet loss \cite{9496127}. However, to examine 6G network performance under stress, we simulate challenging conditions using non-standard link variations, beyond those defined by ITU-T \cite{ITUTY1541} and GSMA \cite{GSMA}, to observe the network's behavior under extreme conditions.
\subsubsection{Scalability} 
While this paper focuses on a specific use case, we recognize the importance of demonstrating its broader applicability. One such application is the mobility metaverse, where each vehicle functions as a sub-network connecting its sensors, with processing occurring in a nearby edge cloud. The proposed topology management system can be adapted to ensure seamless connectivity and low-latency communication for these applications. By dynamically monitoring and predicting changes in network conditions—such as vehicle movement, edge server capacity, or link quality—the system can enable proactive reconfigurations, ensuring reliable data transmission between the vehicle sub-networks and the edge cloud.
\section{Conclusion} \label{sec:conclusion}
In this work, we evaluated a novel topology management system within a dynamic 6G network organized into autonomous sub-networks, focusing on an edge video distribution use case aligned with the latest standards. We introduced an intelligent algorithm for predicting topology changes, capable of optimizing, training, and selecting the most suitable machine learning model for specific scenarios. A cost model was introduced to compare traditional monitoring with our ML-based approach. 
Results show that, for link changes, ANN demonstrates superior accuracy, while XGBoost is more efficient for predicting mobility-based topology changes. Overall, the ML-based approach represents a cost-effective alternative to conventional methods.

Topology change is a particularly interesting topic. Understanding and discovering the impact of these changes on the application layer remains an open area for exploration in future 6G networks, especially given the increasingly dynamic nature of network environments.
\section*{Acknowledgment}
This work was supported by the German Federal Ministry of Education and Research (BMBF) project 6G-ANNA, grant agreement number 16KISK100.
\bibliographystyle{IEEEtran}
\bibliography{iccBib.bib}

\end{document}